\documentclass[twocolumn,floatfix,showpacs]{revtex4}

\usepackage{amsfonts}
\usepackage{amsmath}
\usepackage{amssymb}
\usepackage{mathrsfs}
\usepackage[english,activeacute]{babel}
\usepackage[latin1]{inputenc}
\usepackage{graphicx}
\usepackage{subfigure}
\usepackage{epsfig}
\usepackage{float}

 \usepackage{color}

\begin{document}

\title{\scriptsize{Rev. Mex. F\'{\i}s. 67(3) 443-446 (2021)}\\
---------------------\\
\Large{Factorization method for some inhomogeneous Li\'enard equations}}

\author{O. Cornejo-P\'erez}\email{octavio.cornejo@uaq.mx}
\affiliation{Facultad de Ingenier\'{\i}a, Universidad Aut\'onoma de Quer\'etaro,\\
Centro Universitario Cerro de las Campanas, 76010 Santiago de Quer\'etaro, Mexico}
\author{S.~C. Mancas}\email{mancass@erau.edu}
\affiliation{Department of Mathematics, Embry-Riddle Aeronautical University, Daytona Beach, FL 32114-3900, USA
}%
\author{H.~C. Rosu}\email{hcr@ipicyt.edu.mx}
\affiliation{IPICyT, Instituto Potosino de Investigacion Cientifica y Tecnologica,\\
Camino a la presa San Jos\'e 2055, Col. Lomas 4a Secci\'on, 78216 San Luis Potos\'{\i}, S.L.P., Mexico}%
\author{C.~A. Rico-Olvera}\email{crico24@alumnos.uaq.mx}
\affiliation{Facultad de Ingenier\'{\i}a, Universidad Aut\'onoma de Queretaro,\\
     Centro Universitario Cerro de las Campanas, 76010 Santiago de Queretaro, Mexico}

\pacs{02.30.Hq; 02.30.Ik \hfill  arXiv: 2101.07828}

\begin{abstract}
\noindent We obtain closed-form solutions of several inhomogeneous Li\'enard equations by the factorization method. 
The two factorization conditions involved in the method are turned into a system of first-order differential equations containing the forcing term. In this way, one can find the forcing terms that lead to integrable cases.
Because of the reduction of order feature of factorization, the solutions are simultaneously solutions of first-order differential equations with polynomial nonlinearities. The illustrative examples of Li\'enard solutions obtained in this way generically have rational parts, and consequently display singularities.\\


%
%
\end{abstract}

\maketitle

\section*{1. Introduction}

The exact solutions of nonlinear ordinary differential equations (ODEs) describe the behavior of a great variety of physical, chemical, biological,
and engineering systems. Widespread systems in these vast areas of research can be described by homogeneous Li\'enard equations which have been intensively studied along the years, see, e.g., \cite{LR03} and the recent review \cite{HL16}. On the other hand, the same type of inhomogeneous equations received relatively less attention despite the remarkable leap forward brought by the discovery of an irregular noise, later termed deterministic chaos, in the case of sinusoidally driven triode circuits by van der Pol and van der Mark in 1927 \cite{PM}. Our focus in this short paper is on inhomogeneous Li\'enard type equations of the form
\begin{equation}\label{eq1}
\ddot{u} +G(u)\dot{u} + F(u) = I(t)~,
\end{equation}
where the dot denotes the time derivative, $d/dt$,
$G(u)$ and $F(u)$ are arbitrary, but usually polynomial, functions of $u$, and the forcing term $I(t)$ is an arbitrary continuous function of time.

\medskip

The main goal of the present article is to show how the factorization method developed in \cite{rc,cr,retal} and the factorization conditions thereof
can be used to obtain some integrable inhomogeneous Li\'enard equations for specific forcing terms.
The key point is that the factorization method helps to reduce the inhomogeneous Li\'enard equations to first-order nonlinear equations, such as Abel and Riccati equations, which are presumably easier to solve in some cases. We recall that the reduction to Riccati equations of the linear Schr\"odinger equations has been extensively used in supersymmetric quantum mechanics, and in older factorization methods as reviewed in \cite{dong,mr}.

\section*{2. The nonlinear factorization}\label{s2}

As in \cite{rc,cr,retal}, we consider the factorization of (\ref{eq1}) 
\begin{equation}\label{eq2}
\bigg[ \frac{d}{dt}-f_2(u) \bigg] \bigg[ \frac{d}{dt}-f_1(u)\bigg]u=I(t)
\end{equation}
under the conditions 
\begin{eqnarray}
f_2+\frac{d(f_1u)}{du}&=-G(u)\label{eq3}\\
f_1 f_2 u &=F(u)~,\label{eq3bis}
\end{eqnarray}
adding the scheme proposed 
in \cite{wang1}, where one assumes $[d/dt-f_1(u)]u=\Omega(t)$.
This yields the following coupled ODEs for~(\ref{eq2}),
\begin{eqnarray}
&\dot{\Omega}-f_2(u)\Omega =I(t)\label{eq4}\\
&\dot{u}-f_1(u)u=\Omega(t)~, \label{eq5}
\end{eqnarray}
which we further simplify by taking the second factorizing function as a constant, $f_2=a_2\equiv const.$,
\begin{eqnarray}
\dot{\Omega}-a_2\Omega &=I(t)\label{eq6}\\
\dot{u}-f_1(u)u &=\Omega(t)~.\label{eq7}
\end{eqnarray}
Besides, using the constant function $f_2$, conditions~(\ref{eq3}) and ~(\ref{eq3bis}) imply a relationship between functions $F$ and $G$ given by
\begin{equation}\label{eq7bis}
F(u)=-a_2\left(c_2+a_2u+\int^u G(u) ~du \right)~,
\end{equation}
where $c_2$ stands for the integration constant, or equivalently
\begin{equation}\label{eq7tris} 
G(u)=-\left(\frac{1}{a_2}\frac{dF}{du}+a_2\right)~.
\end{equation}
Denoting ${\cal I}(t)=\int_0^t e^{-a_2 t}I(t)dt$, the  solution to (\ref{eq6}) is
\begin{equation}\label{eq10}
\Omega(t)= e^{a_2 t} \big[ c_1 + {\cal I}(t)\big]~, 
\end{equation}
where $c_1$ is an integration constant given by $c_1=\Omega(0)$. This allows to rewrite (\ref{eq7}) in the form
\begin{equation}
\dot{u}   
=\frac{1}{a_2}F(u)+ e^{a_2 t} \big[ c_1 + {\cal I}(t)\big]~, \label{eq8}
\end{equation}
whose general solution is also the solution of the Li\'enard equation~(\ref{eq1}), while further particular solutions can be obtained by setting $c_1=0$.

\medskip

Viceversa, one can say that (\ref{eq8}) 
is a first-order nonlinear reduction of forced Li\'enard equations of the form
\begin{equation}\label{eq9}
\ddot{u}-\left(\frac{1}{a_2}\frac{dF}{du}+a_2\right)\dot{u} + F(u)=I(t)~.
\end{equation}
Thus, integrable cases of (\ref{eq8}) can provide Li\'enard solutions in closed form.
Since among the most encountered forced Li\'enard equations are these having $F(u)$ in the form of cubic and quadratic polynomials, in the rest of the paper, we address the applications of this solution method to some cases of these types.

\section*{3. The inhomogeneous Duffing-van der Pol oscillator}\label{s3}

We choose the particular cubic case $F(u)=Au+Cu^3$ because it corresponds to the forced Duffing-van der Pol oscillator \cite{chandra1}
\begin{equation}
\ddot{u} - [ (a_2+A/a_2) + 3(C/a_2) u^2 ]\dot{u} + Au + Cu^3=I(t)~.\label{duf-1}
\end{equation}
This equation admits the factorization
\begin{equation}
\bigg[\frac{d}{dt}-a_2\bigg]\bigg[\frac{d}{dt}-(\alpha +\gamma u^2)\bigg]u=I(t)~, \label{duf-2}
\end{equation}
where $\alpha=A/a_2$ and $\gamma=C/a_2$.

The corresponding first-order equation is the Abel equation 
\begin{equation}
\dot{u}= \gamma u^3 + \alpha u+ \Omega(t)~. \label{duf-3}
\end{equation}
The change of variables 
\begin{eqnarray}
 u= y e^{\alpha t}~, \quad\quad  x= \frac{\gamma}{2\alpha}e^{2\alpha t}~,  
\end{eqnarray}\label{duf-4}
turns ~(\ref{duf-3}) into the normal form
\begin{equation}\label{norm2}
\frac{dy}{dx}=y^3 +{\cal N}(x)~,
\end{equation}
with invariant
\begin{equation}\label{Ncubic}
{\cal N}(x) = \frac{1}{\gamma}e^{(a_2-3\alpha) t(x)} \big[ c_1 + {\cal I}(t(x))\big]~.
\end{equation}
Unfortunately, this formula shows that inhomogeneous Abel equations in this category are not integrable by separation of variables because ${\cal N}(x)$ cannot be made constant as required by this type of integrability.
Only in the force-free particular case $I(t)=0$, the invariant can be reduced to the constant
\begin{equation}\label{N0}
{\cal N}_0=\frac{c_1}{\gamma}~. 
\end{equation}
By separation of variables, the solution 
is given by the implicit relation
\begin{eqnarray}\label{solnorm}
&\ln\left[\frac{{(\sqrt[3]{\cal N}_0+y)^2}}{{\cal N}_0^{2/3}-\sqrt[3]{\cal N}_0y+y^2}\right]-2 \sqrt{3} \tan ^{-1}\left[\frac{1-\frac{2}{\sqrt[3]{\cal N}_0}y}{\sqrt{3}}\right] =\nonumber \\
&6{\cal N}_0^{2/3}(x+c_2)~.
\end{eqnarray}
This solution has been obtained previously in \cite{chandra1}.

\section*{4. Quadratic inhomogeneous Li\'enard equations}\label{s4}

\bigskip

If we set $F(u)=Au+Bu^2$, 
then the first order equivalent equation is the  Riccati equation
\begin{equation}\label{eq11-1}
\dot{u}= \beta u^2 + \alpha u+\Omega(t)~,  \qquad \beta=B/a_2~.
\end{equation}
Equation~(\ref{eq11-1}) can be transformed into the normal form  \cite{saraf} 
\begin{equation}\label{norm3}
\dot { z } =  { z }^{ 2 } + {\cal N}(t)~,
\end{equation}
where
\begin{eqnarray}
z(t)=\beta u(t)+\frac{\alpha}{2}, \quad {\cal N}(t)=\beta \Omega(t)-\frac{\alpha^2}{4}~.
\end{eqnarray}
For integrable cases of separable type, one should have ${\cal N}(t)$ as an arbitrary real constant that we choose $p^2/4$ implying  $\Omega(t)=\left(p^2+\alpha^2\right)/4\beta$ also a constant, as well as a constant driving force
\begin{equation}\label{Iconst1}
I(t)=-\frac{a_2}{\beta}\left(\frac{p^2+\alpha^2}{4}\right)~.
\end{equation}
In this simple case, we obtain a Li\'enard solution of ~(\ref{eq9}) of the form 
\begin{equation}\label{soltan}
u(t)=-\frac{\alpha}{2\beta}\bigg[1 -\frac{p}{\alpha} \tan \left(\frac{p}{2} (t+c_2)\right)\bigg]~.
\end{equation}

\subsection*{4.1 Linear polynomial source term}

\medskip

 After the constant driving, it is orderly to consider the source term as the linear polynomial $I(t)= t + \delta$, where $\delta$ is an arbitrary constant. We set $a_2=1$ and $c_1=0$, and we obtain the Riccati equation
\begin{equation}\label{eq2-4}
\dot { u } = \beta { u }^{ 2 }+ \alpha u -(t+\tilde{\delta})~,\qquad \tilde{\delta}=\delta+1
\end{equation}
with solution given by
\begin{equation}\label{eq2-5} 
 u(t) = - \frac{\alpha}{2\beta}\bigg[1 +\frac{\beta^{\frac{1}{3}}}{\alpha}
 \frac{k_2 Ai^{\prime}(\tilde t)+ Bi^{\prime}(\tilde t )}
 {k_2 Ai(\tilde t) +Bi(\tilde t)}\bigg] ~,
\end{equation}
where $\tilde t=\beta^{1/3}[\alpha^2/4\beta +(t+\tilde{\delta}]$, the prime denotes the $\tilde t$ derivative, and  $k_2$ is an integration constant.
However, the presence of the rational term in Airy functions turns singular such kinds of Li\'enard solutions.

\subsection*{4.2 Quadratic polynomial source term}

\medskip

Let the source term be the quadratic polynomial of type
$I(t)=a_2\beta t^{2}+ (a_2\alpha -2\beta)t -(a_2+\alpha)$.
According to Eqs.~(\ref{eq10}) and (\ref{eq11-1}), and by setting $c_1=0$, we have the Riccati equation
\begin{equation}\label{eq2-1}
    \dot { u } = \beta { u }^{ 2 }+ \alpha u - \beta t^2 -\alpha t + 1~.
\end{equation}
This equation has the particular solution $u(t)=t$, while the general solution is given by
\begin{equation} \label{eq2-2}
 u(t)=t-\frac{e^{t(\alpha+\beta t)}}{ k_1 \beta + e^{t(\alpha+\beta t)} \sqrt{\beta} {\cal F}\left( \frac{\alpha+2\beta t}{2\sqrt{\beta}}\right)}~,
\end{equation}
where ${\cal F}(x)=e^{-x^2}\int_0^x{e^{y^2} dy}$ is the Dawson integral, and $k_1$ is an integration constant. Again, because of the rational term this
solution is singular at $-e^{t(\alpha+\beta t)} \beta^{-1/2} {\cal F}\left( \frac{\alpha+2\beta t}{2\sqrt{\beta}}\right)=k_1$.

\subsection*{4.3 Exponential source term}

\medskip

For the source term of exponential form, $I(t)= \kappa e^{\lambda t}$, and for $c_1=0$, the Riccati equation is
\begin{equation}\label{eq2-9}
    \dot{ u } = \beta { u }^{ 2 }+ \alpha u + \frac{\kappa}{\lambda- a_2} e^{\lambda t}~. 
\end{equation}
The solution is given by
\begin{equation}\label{eq2-10}
 u(t)  = \frac{\alpha}{2\beta}\bigg[\frac{\lambda}{\alpha} \frac{2 k_4 \Gamma(1-\frac{\alpha}{\lambda})\tilde t J_{1 -\frac{\alpha}{\lambda}}(2\tilde t)
 -\tilde t^{\frac{\alpha}{\lambda}} \bar{F}(\tilde t)}
 { k_4 \Gamma(1-\frac{\alpha}{\lambda})J_{-\frac{\alpha}{\lambda}}(2\tilde t) + \Gamma(1+\frac{\alpha}{\lambda})J_{\frac{\alpha}{\lambda}}(2\tilde t)}\bigg]~,
\end{equation}
 where $k_4$ is an integration constant, $\tilde t =2\frac{\sqrt{\kappa\beta}}{\lambda\sqrt{\lambda- a_2}}e^{\lambda t/2}$,
 and $\bar{F}(\tilde t)$ is the following combination of hypergeometric functions
 \begin{eqnarray}
 &\bar{F}(\tilde t)  = 
 {\tilde t}^2\, {}_0F_{1}\left(;2+\frac{\alpha}{\lambda};-\tilde t^2\right)+\frac{\alpha}{\lambda} \, {}_0F_{1}\left(;1+\frac{\alpha}{\lambda};-\tilde t^2\right)\nonumber\\
 &+ \, {}_0F_{1}\left(;\frac{\alpha}{\lambda};-\tilde t^2\right)~. \nonumber
 \end{eqnarray}

For $k_4=0$, we have the simpler solution
\begin{equation}\label{simpb} 
u(t)=-\frac{\alpha}{2\beta}\bigg[1+\frac{\lambda}{\alpha}\,\frac{\tilde t^2\, _0{F}_1\left(;2+\frac{\alpha}{\lambda };-\tilde t^2\right)+\,
_0{F}_1\left(;\frac{\alpha}{\lambda};-\tilde t^2\right)}{ _0{F}_1\left(;1+\frac{\alpha}{\lambda};-\tilde t^2\right)}\bigg]~.
\end{equation}

The case corresponding to $\alpha=-1$ simplifies further to
\begin{equation}\label{tan}
u(t)=\frac{e^t}{\sqrt{\beta }} \tan \left[\sqrt{\beta} (e^t+k_4)\right].
\end{equation}

\subsection*{4.4 Back to the constant source case}

We return to the constant source term case since we wish to point out the interesting feature that it is more general than the exponential case.
Indeed, let us take the source term as $I(t)= \epsilon$, an arbitrary constant, and $a_2=1$. This leads to the Riccati equation
\begin{equation}\label{eq2-6}
    \dot { u } = \beta { u }^{ 2 }+ \alpha u + c_1 e^{t} - \epsilon~,
\end{equation}
which is similar to the Riccati equation for the exponential case unless for $\epsilon$.
The general solution of (\ref{eq2-6}) is a rational expression in Bessel functions given by
\begin{eqnarray} \label{eq2-7}
& u(t) =\frac{\alpha}{2\beta}\bigg[\frac{m}{\alpha}\frac{k_3(\alpha-m)\Gamma (-m) J_{-m}(\tilde{t})
 -(\alpha+m)\Gamma (m) J_{m}(\tilde{t})}
 {k_3\Gamma (1-m) J_{-m}(\tilde{t})+\Gamma (1+m) J_m(\tilde{t})}\bigg]\nonumber \\
&+\frac{\alpha}{2\beta}\bigg[ \frac{\tilde{t}}{\alpha}\frac{k_3\Gamma (1-m) J_{1-m}(\tilde{t})+m\Gamma (m) J_{1+m}(\tilde{t}))}
 {k_3\Gamma (1-m) J_{-m}(\tilde{t})+\Gamma (1+m) J_m(\tilde{t})}\bigg]~,
\end{eqnarray}
where $m=\sqrt{\alpha ^2+4 \beta  \epsilon}$, $\tilde{t}=2\sqrt{\beta c_1} e^{t/2}$, and $k_3$, an integration constant.
It displays singularities at the zeros of its denominators.

\medskip

\noindent When $k_3=0$, this solution takes the simpler form
\begin{equation}\label{eqz2} 
u(t)=-\frac{\alpha}{2\beta}\bigg[\left(1+\frac{m}{\alpha}\right)-\frac{\tilde{t}}{\alpha}\,\frac{J_{m+1}(\tilde{t})}{J_m(\tilde{t})}\bigg]~.
\end{equation}

Notice that in the particular case of $c_1=0$, the exponential scaling of time is annihilated and the Riccati equation is of constant coefficients having the well known regular kink solution
\begin{equation} \label{eq29} 
u(t)=-\frac{\alpha}{2\beta}\bigg[1+\frac{m}{\alpha} \tanh \left(\frac{m}{2} (t+k_3)\right)\bigg]~,
\end{equation}
which is also a Li\'enard kink. If in the expression for the parameter $m$ we substitute $\epsilon$ by (\ref{Iconst1}) for $a_2=1$, we obtain $m=ip$, and
(\ref{eq29}) becomes the solution (\ref{soltan}).

\section*{5. Conclusion}\label{s5}

The nonlinear factorization method developed in \cite{rc,cr,retal,wang1} has been used to obtain closed-form solutions of certain types of inhomogeneous Li\'enard equations. The conditions imposed upon the nonlinear coefficients of the equations by the factorization method and the insertion of the forcing term in the factorization scheme act as designing tools of specific forms of the forcing terms to generate integrable cases by these means. The illustrative examples have been chosen from the class of polynomial (up to cubic) and exponential forcing terms similar to a recent study of inhomogeneous Airy equations \cite{Dunster}. However, the obtained Li\'enard solutions have rational parts which make them prone to the presence of singularities. The only regular solutions that we have obtained by employing this simple factorization method are the usual tanh kinks.
Finally, the scheme presented here is bounded to constant factorization functions $f_2$, since only in this case equation~(\ref{eq4}) can be turned into the linear equation~(\ref{eq6}) in the independent variable $t$.


%


\end{document}